\documentclass[sigconf,screen, authorversion]{acmart}
\AtBeginDocument{%
  \providecommand\BibTeX{{%
    \normalfont B\kern-0.5em{\scshape i\kern-0.25em b}\kern-0.8em\TeX}}}

\setcopyright{rightsretained}
\copyrightyear{2023}
\acmYear{2023}
\acmDOI{XXXXXXX.XXXXXXX}

\acmConference[CHI '23]{ACM CHI Conference on Human Factors in Computing Systems}{April 23-28, 2023}{Hamburg, Germany}
\acmBooktitle{CHI '23 Workshop on Future of Computational Approaches for Understanding and Adapting User Interfaces: ACM CHI Conference on Human Factors in Computing Systems,
 April 23, 2023, Hamburg, Germany} 
\usepackage{tikz}
\usetikzlibrary{shapes.geometric, arrows}

\tikzstyle{arrow} = [thick,->,>=stealth]

\usepackage{caption}
\usepackage{subcaption}

\usepackage{amsmath}
\usepackage{lscape}
\usepackage{enumitem}

\usepackage{tabularx}
\newcolumntype{P}[1]{>{\centering\arraybackslash}p{#1}}
\newcolumntype{M}[1]{>{\centering\arraybackslash}m{#1}}
\setlist[itemize]{leftmargin=*}
\setlist[enumerate]{leftmargin=*,label=\arabic*.}
\begin{document}

\title{Mixed Reality UI Adaptations With Inaccurate and Incomplete Objectives}

\author{Christoph Albert Johns}
\affiliation{%
  \institution{Aarhus University}
  \country{Denmark}
}
\email{cajohns@cs.au.dk}

\author{João Marcelo Evangelista Belo}
\affiliation{%
  \institution{Aarhus University}
  \country{Denmark}
}
\email{joaobelo@cs.au.dk}

\renewcommand{\shortauthors}{Johns and Belo}

%% article (max. 150 words).
\begin{abstract}
This position paper outlines a new approach to adapting 3D user interface (UI) layouts given the complex nature of end-user preferences.
Current optimization techniques, which mainly rely on weighted sum methods, can be inflexible and result in unsatisfactory adaptations.
We propose using multi-objective optimization and interactive preference elicitation to provide semi-automated, flexible, and effective adaptations of 3D UIs.
Our approach is demonstrated using an example of single-element 3D layout adaptation with ergonomic objectives.
Future work is needed to address questions around the presentation and selection of optimal solutions, the impact on cognitive load, and the integration of preference learning. 
We conclude that, to make adaptive 3D UIs truly effective, we must acknowledge the limitations of our optimization objectives and techniques and emphasize the importance of user control.
\end{abstract}


\begin{CCSXML}
<ccs2012>
   <concept>
       <concept_id>10003120.10003138.10003140</concept_id>
       <concept_desc>Human-centered computing~Ubiquitous and mobile computing systems and tools</concept_desc>
       <concept_significance>500</concept_significance>
       </concept>
 </ccs2012>
\end{CCSXML}

\ccsdesc[500]{Human-centered computing~Ubiquitous and mobile computing systems and tools}

\keywords{multi-objective optimization, Pareto frontier, UI adaptation, mixed reality}

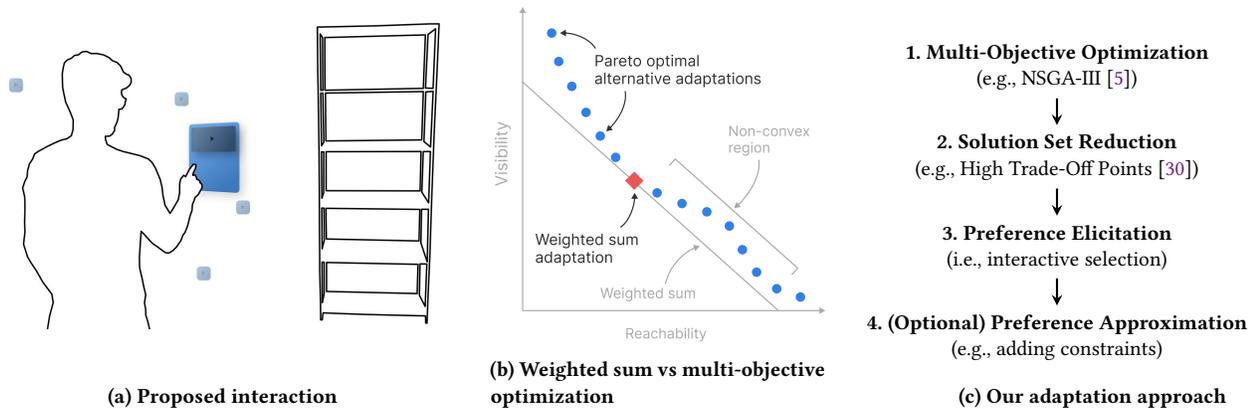
\begin{teaserfigure}
    \centering
    \begin{subfigure}[b]{0.35\textwidth}
         \centering
          \includegraphics[width=\linewidth]{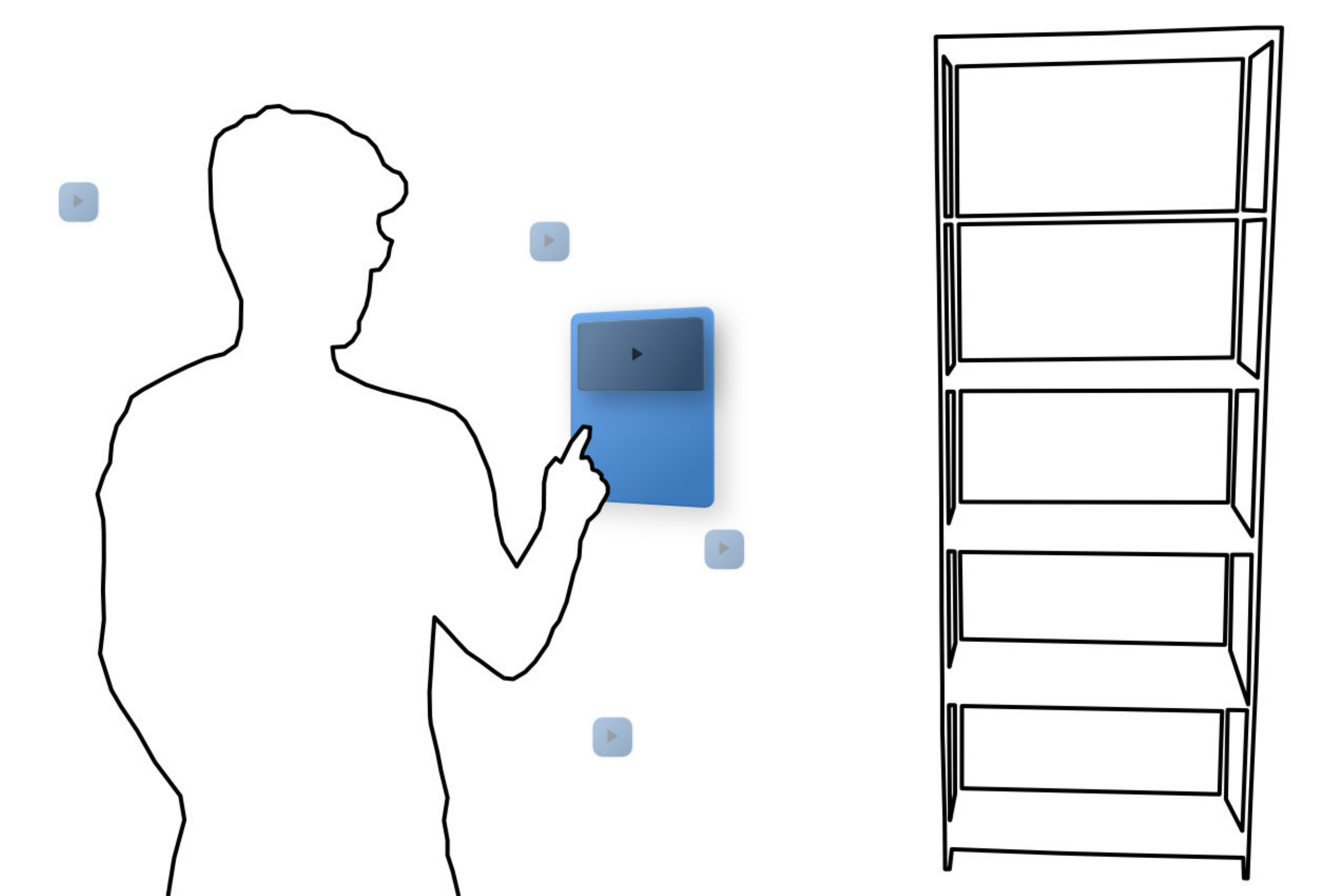}
          \vspace{4pt}
          \caption{Proposed interaction}
          \label{fig:imagined_interaction}
          \Description{3D vector illustration of a person standing in front of a bookshelf interacting with a floating user interface panel via touch. Around the floating panel, four additional small icons are distributed in the 3D scene and not attended to by the person.}
    \end{subfigure}
    \hfill
    \begin{subfigure}[b]{0.25\textwidth}
         \centering
         \includegraphics[width=\linewidth]{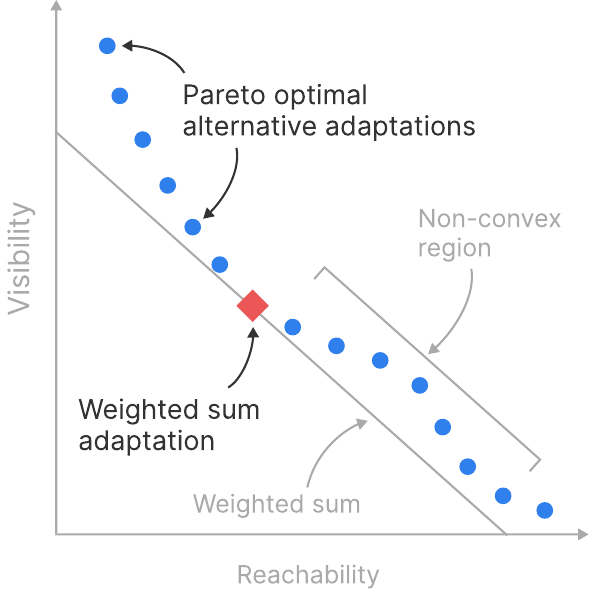}
        \caption{Weighted sum vs multi-objective optimization}
        \Description{A scatter plot with two axes. The x-axis is labeled 'Reachability'. The y-axis is labeled 'Visibility'. A line at 45 degrees going through the plot is labeled 'Weighted sum' and intersects with the y-axis at a high point and with the x-axis at a high value. Fifteen points are highlighted in the plot. A point on the 45-degree line is labeled 'Weighted sum adaptation' and colored red. The other points form a border with convex and non-convex regions above the 45-degree line and are colored blue. The blue points are labeled 'Pareto optimal alternative adaptations'. The points on the non-convex region are labeled 'Non-convex region'.}
        \label{fig:pareto_front}
     \end{subfigure}
     \hfill
     \begin{subfigure}[b]{0.35\textwidth}
         \begin{tikzpicture}[node distance=1.2cm]
            \node[font=\small] (opt) [align=center] {\textbf{1. Multi-Objective Optimization} \\(e.g., NSGA-III~\cite{deb_investigating_2019})};
            \node[font=\small] (red) [below of=opt, align=center] {\textbf{2. Solution Set Reduction} \\(e.g., High Trade-Off Points~\cite{rachmawati_multiobjective_2009})};
            \node[font=\small] (eli) [below of=red, align=center] {\textbf{3. Preference Elicitation} \\(i.e., interactive selection)};
            \node[font=\small] (appr) [below of=eli, align=center] {\textbf{4. (Optional) Preference Approximation} \\(e.g., adding constraints)};
            \draw [arrow] (opt) -- (red);
            \draw [arrow] (red) -- (eli);
            \draw [arrow] (eli) -- (appr);
        \end{tikzpicture}
        \caption{Our adaptation approach}
        \label{fig:approach}
        \Description{Flowchart with four elements linearly connected by flow links. The start state is ‘1. Multi-Objective Optimization (i.e., NSGA-III)’. The second and third state are '2. Solution Set Reduction (i.e., High Trade-Off Points)' and '3. Preference Elicitation (i.e., Interactive selection)'. The end state is ‘4. (Optional) Preference Approximation (e.g., adding constraints)’.}
     \end{subfigure}
     \caption{We present an approach for 3D user interface layout adaptation utilizing online multi-objective optimization. Multiple Pareto optimal potential adaptations representing trade-offs between pre-defined objectives are generated and presented to the user for selection. (a) Illustration of a mixed reality assembly application implementing our proposed approach (here, to place instructions on assembling furniture). (b) Illustrative comparison of an adaptation found using weighted sum optimization (red) and our approach (blue). (c) Overview of our proposed method for optimization and preference elicitation.}
     \label{fig:teaser}
\end{teaserfigure}

\maketitle

\section{Introduction}
\label{sec:introduction}

Adaptive user interfaces automatically configure their appearance and behavior to fit the current context of use \cite{lindlbauer_context-aware_2019}.
This is particularly important --- yet quite challenging --- in the domain of mixed reality (MR) where users interact with complex environments that blend physical and virtual worlds. Here various contextual factors such as lighting conditions, user pose, and the presence of other people influence the overall usability of the MR interface (see \autoref{fig:imagined_interaction}; \cite{lindlbauer_context-aware_2019, cheng_semanticadapt_2021}).
One popular approach to the adaptation of 3D user interfaces (UIs) is optimization (e.g., \cite{belo_auit_2022, cheng_semanticadapt_2021, lindlbauer_context-aware_2019, fender_optispace_2018}).
Here, the UI is configured by optimizing one or multiple objective functions that are assumed to relate to end-user satisfaction or preferences.
A 3D UI layout may, for example, be configured to minimize the ergonomic stress put on the user's body \cite{evangelista_belo_xrgonomics_2021} or to maximize the relevance of the displayed information \cite{lindlbauer_context-aware_2019}.
End-user satisfaction and preferences are, however, latent and insufficiently understood in the domain of 3D user interfaces and of MR applications in particular, as the scale and contextual complexity of MR environments pose particular challenges for computational modelling.
This issue is exacerbated by the fact that context itself may only be partially captured or approximated by MR systems \cite{lu_exploring_2022}.
As a result, most attempts to optimize MR UIs use indirect, incomplete and approximate objectives dependent on a few measured contextual factors that aim to capture some aspect of end-user satisfaction or preferences (e.g., related to ergonomics as described above) but cannot make any guarantees about the quality of the resulting adaptation as perceived by the end-user.

Despite these limitations, many adaptive systems for MR UIs propose fully automated solutions and provide few means for end-users to meaningfully influence the adaptation process.
Consequently, end-users are often left with the choice between fully automated adaptations that may not be satisfactory and manual adaptations that are time-consuming and error-prone.
Typically, these fully automated adaptation systems employ a static linear combination of their objective functions to act as a global optimization criterion and adapt the UI by minimizing or maximizing this overall cost or utility function (see \autoref{fig:pareto_front}; e.g., \cite{belo_auit_2022, cheng_semanticadapt_2021, lindlbauer_context-aware_2019, fender_optispace_2018, xiao_supporting_2017, nuernberger_snaptoreality_2016, gal_flare_2014}).
This assumes that the set of objectives and their weighted combination --- with weights either specified by the designer or by the end-user --- are a sufficient representation of the end-user's true preference function.
% Here I think we should start by arguing about how static weights cannot encapsulate variable preferences
Such an approach typically employs static weights, relying on the assumption that user preferences towards these adaptation objectives remain constant over time and are the same for the whole contextual input space. Furthermore, it does not acknowledge the limitations of the underlying approximate objective functions nor of the optimization technique itself --- linear global criteria are, for example, unable to capture solutions in non-convex regions of the Pareto frontier (cf. \cite{marler_survey_2004}) --- and may, consequently, lead to unsatisfying adaptations.
% This are great questions, but I'm unsure how effectively our current approach fits some edge cases (e.g. have sufficiently accurate objectives to optimize). In this particular case the reader might think about unspecified objectives. We discussed this in person, but the pipeline we propose here does not try to learn new objectives from the input space, a point that could be confused with this question.
The question is then: How can we create adaptive 3D user interfaces today if we don't have sufficiently accurate objectives to optimize? How can we adapt to end-user preferences that are latent, complex, and costly to query? Given that preferences and contextual factors are likely incomplete and only approximated?

In this paper, we present a novel approach to the adaptation of MR UIs that acknowledges the limitations of approximate objective functions and provides end-users with the opportunity to meaningfully influence the adaptation process (see \autoref{fig:approach}).
This approach is based on online multi-objective optimization and interactive \textit{a posteriori} preference elicitation.
It allows end-users to select a base adaptation from a set of Pareto optimal candidates and to manually correct the adaptation in the 3D environment.
Furthermore, our approach provides the opportunity for full automation and continuous preference learning if desired.
Overall, we attempt to expose more of the design space of potential adaptations while reducing the set of presented solutions to those that are \textit{optimal} or \textit{efficient} according to some pre-defined criterion, improving user control compared to fully automated adaptation.
%while reducing the cognitive load compared to fully manual adjustment. --- (Joao) This is a bit of a stretch right now / at least I don't think we plan to assess this claim at any point in the future. We can chat about it
We illustrate our approach using the example of single-element layout adaptation in a 3D environment based on simple ergonomic metrics inspired by XRgonomics~\cite{evangelista_belo_xrgonomics_2021} and RULA~\cite{mcatamney_rula_1993}.
By comparing our approach to the state-of-the-art weighted sum approach, we show that it finds a variety of trade-offs between objectives, some of which are more likely to be preferred than the adaptations suggested by the state-of-the-art weighted sum methods.
We conclude by discussing the limitations of our approach and the opportunities for future work.

\section{Related Work}
\label{sec:related_work}

Our work builds upon research on the adaptation of UI layouts in MR applications using global criterion optimization and on previous efforts in multi-objective optimization of UIs outside the domain of MR.
Global criterion optimization for adaptive MR refers to the real-time adaptation of MR interfaces using online optimization of a single scalarized utility or cost function.

\subsection{Optimization of User Interfaces}

Two important milestones for applying optimization techniques in UI design were the representation of design factors as decision variables in problem formulations and the modeling of design heuristics and psychological models in cost functions~\cite[sec.~4.2]{oulasvirta2018computational}.
Pioneering work proposed novel objective functions and used existing optimization techniques to optimize keyboard layouts~\cite{burkard1977entwurf, light1993designing, zhai2002performance}, UI layouts~\cite{gajos_supple_2004, gajos_preference_2005, gajos_automatically_2010} and menu designs~\cite{liu2002applying, matsui2008genetic, goubko2010automated, gilles2013menuoptimizer}.
Since then, multi-objective optimization approaches have been applied to the design and offline adaptation of, for example, text entry methods (e.g., ~\cite{gong_wristext_2018, bi_ijqwerty_2016, smith_optimizing_2015, oulasvirta_improving_2013, dunlop_multidimensional_2012, sridhar_investigating_2015}) and graphical UIs (e.g., \cite{todi2016}).
Recently, alternative multi-objective optimization methods have been utilized in the domain of keyboard layout optimization~\cite{lee_fingertext_2021}, computational design of physical input devices~\cite{liao_computational_2021} and parameter search for 3D touch interaction~\cite{chan_investigating_2022}.
These efforts employ more complex optimization methods (i.e., genetic algorithms~\cite{lee_fingertext_2021} and Bayesian multi-objective optimization~\cite{liao_computational_2021, chan_investigating_2022}) to overcome the limitations of weighted sum approaches and find optimal configurations before deployment.
% Summary of review
Overall, multi-objective optimization of UIs has been dominated by weighted sum-based methods used at design time.
As the recent efforts described above demonstrate, however, alternative approaches exist that aim to address some of the shortcomings of weighted sum-based optimization for UI adaptation.
We contribute to this growing area of research by exploring the use of state-of-the-art multi-objective optimization techniques for online layout adaptation of MR UIs.

\subsection{Global Criterion Optimization for Adaptive Mixed Reality}

Several works have applied optimization approaches to adaptive layout problems in MR.
These previous efforts are commonly concerned with optimizing UI elements' visibility (e.g., \cite{fender_optispace_2018}) and spatial coherence (e.g., \cite{nuernberger_snaptoreality_2016,cheng_semanticadapt_2021}), but have also considered other objectives such as ergonomics \cite{evangelista_belo_xrgonomics_2021} and cognitive load \cite{lindlbauer_context-aware_2019}.
Commonly, these optimization-based adaptation approaches treat layout adaptation as a single-objective optimization problem, combining multiple objectives for multiple elements into a single cost to be minimized or a single utility score to be maximized and returning a single solution.
Most frequently, this global criterion takes the form of a static weighted sum (e.g., \cite{belo_auit_2022, cheng_semanticadapt_2021, lindlbauer_context-aware_2019, fender_optispace_2018, xiao_supporting_2017, nuernberger_snaptoreality_2016, gal_flare_2014}), which can be optimized using commercial solvers for linear programs (e.g., \cite{lindlbauer_context-aware_2019, cheng_semanticadapt_2021}) or using approximate techniques like simulated annealing (e.g., \cite{belo_auit_2022}).
In contrast to these previous efforts, our approach treats layout adaptation as an online multi-objective optimization problem that returns a set of optimal adaptations.
Consequently, our proposed approach is not restricted to the same requirements as global criterion methods, for example regarding the formulation of static weights or normalization~\cite{marler_survey_2004}, and can explore novel interfaces that provide more flexibility to users, allowing them to choose from multiple solutions along the Pareto frontier.
Notably, this does not implicate that adaptations cannot be fully automatic.
Even with a Pareto set of solutions, the system can take initiative by picking one solution and showing its alternatives upon request.

\section{Approach}
\label{sec:approach}

We propose a novel technique for real-time UI adaptation in MR applications that uses multi-objective optimization to produce a set of Pareto optimal adaptations that users can choose from, providing a balance between fully automated and fully manual adaptation.
The approach consists of four steps: Multi-Objective Optimization, Solution Set Reduction, Preference Elicitation and (optionally) Preference Approximation (see \autoref{fig:approach}).

% Multi-Objective Optimization
\textit{Multi-Objective Optimization.} Multi-objective optimization refers to finding the optimal solutions for a set of objective functions given some inequality and equality constraints \cite{marler_survey_2004}, for example, minimizing the expected neck and shoulder strain caused by a UI element's placement as constrained by the user's reach.
Typically, the solution to a multi-objective optimization problem is a set of optimal trade-offs between conflicting objectives called the Pareto front \cite{marler_survey_2004}.
While many multi-objective optimization methods exist that can generate Pareto optimal solutions (see, for example, \cite{marler_survey_2004} for an overview), we choose to employ a recent evolutionary algorithm called non-dominated sorting genetic algorithm III (NSGA-III)~\cite{deb_investigating_2019} to demonstrate our approach.
The result of the multi-objective optimization step is a set of Pareto optimal solutions, which we refer to as potential adaptations.

% Solution Set Reduction
\textit{Solution Set Reduction.} Taking the set of Pareto optimal potential adaptations as input, we reduce the set of solutions to a limited set of qualitatively different adaptations.
Literature on multi-objective optimization and multi-criteria decision-making provides a variety of methods to achieve this goal (see, for example, \cite{thakkar_introduction_2021} for an overview).
For our demonstration, we employ a metric described in \cite{rachmawati_multiobjective_2009} to identify efficient optimal trade-off points along convex buldges in the Pareto front.

% Preference Elicitation
\textit{Preference Elicitation.} During Preference Elicitation, we present the reduced set of Pareto optimal solutions to the user.
The user, then, makes their preferences for certain layouts explicit by choosing which proposed adaptations to interact with.
We refer to this activity as preference elicitation.
Depending on the specific adaptation and application context, many methods of preference elicitation may be appropriate.
In our illustrative example, we display the potential UI adaptations using a semi-transparent duplicate of the original UI, which the user can touch to select the adaptation they prefer.
Formally, this step can be omitted to produce a fully automated adaptation technique.
However, recent work has shown, that, in the presence of inaccuracy such as that caused by approximated preference criteria, end-users prefer to give up some level of automation in exchange for increased control over the adaptation process \cite{lu_exploring_2022}.

% (Optional) Preference Approximation
\textit{(Optional) Preference Approximation.} The final and optional step in our proposed approach is Preference Approximation.
Interpreting the user's selection from a set of presented solutions as an explicit statement of preference, this information can be used to inform an approximation of the user's latent preference function.
This approximation can then, in turn, be used to guide future runs of the multi-objective optimization, for example by adding further constraints (see, for instance, \cite{gajos_automatically_2010}).
While our implementation does not take advantage of the user selection due to its effects on system complexity, a continuously updating preference function could be used to effectively improve later adaptations.

\section{Explorative Simulation}
\label{sec:explorative_simulation}

\begin{figure*}[htb]
    \centering
    \begin{subfigure}[b]{0.15\textwidth}
         \centering
          \includegraphics[width=\linewidth]{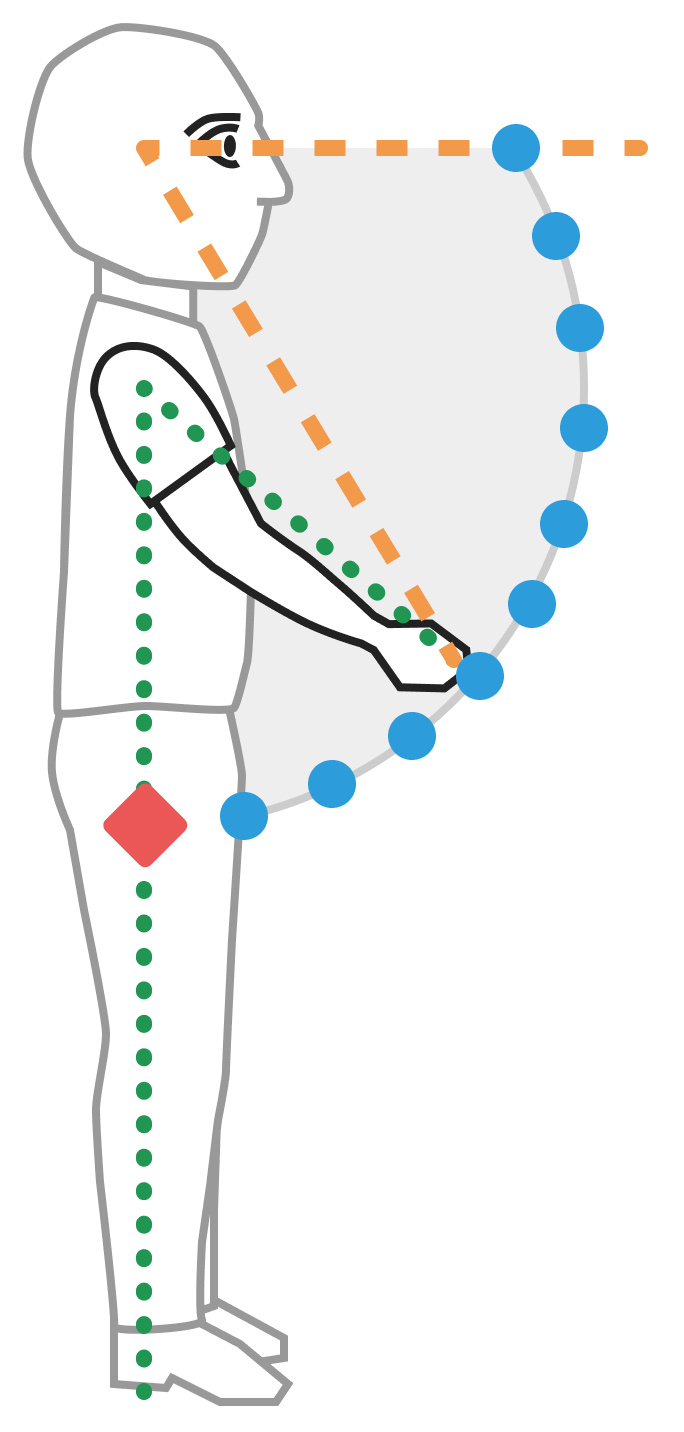}
          \caption{Illustration}
          \label{fig:illustration}
          \Description{2D vector illustration of a person in a full-body profile with an outstretched arm.}
    \end{subfigure}
    \hfill
    \begin{subfigure}[b]{0.4\textwidth}
         \centering
          \includegraphics[width=\linewidth]{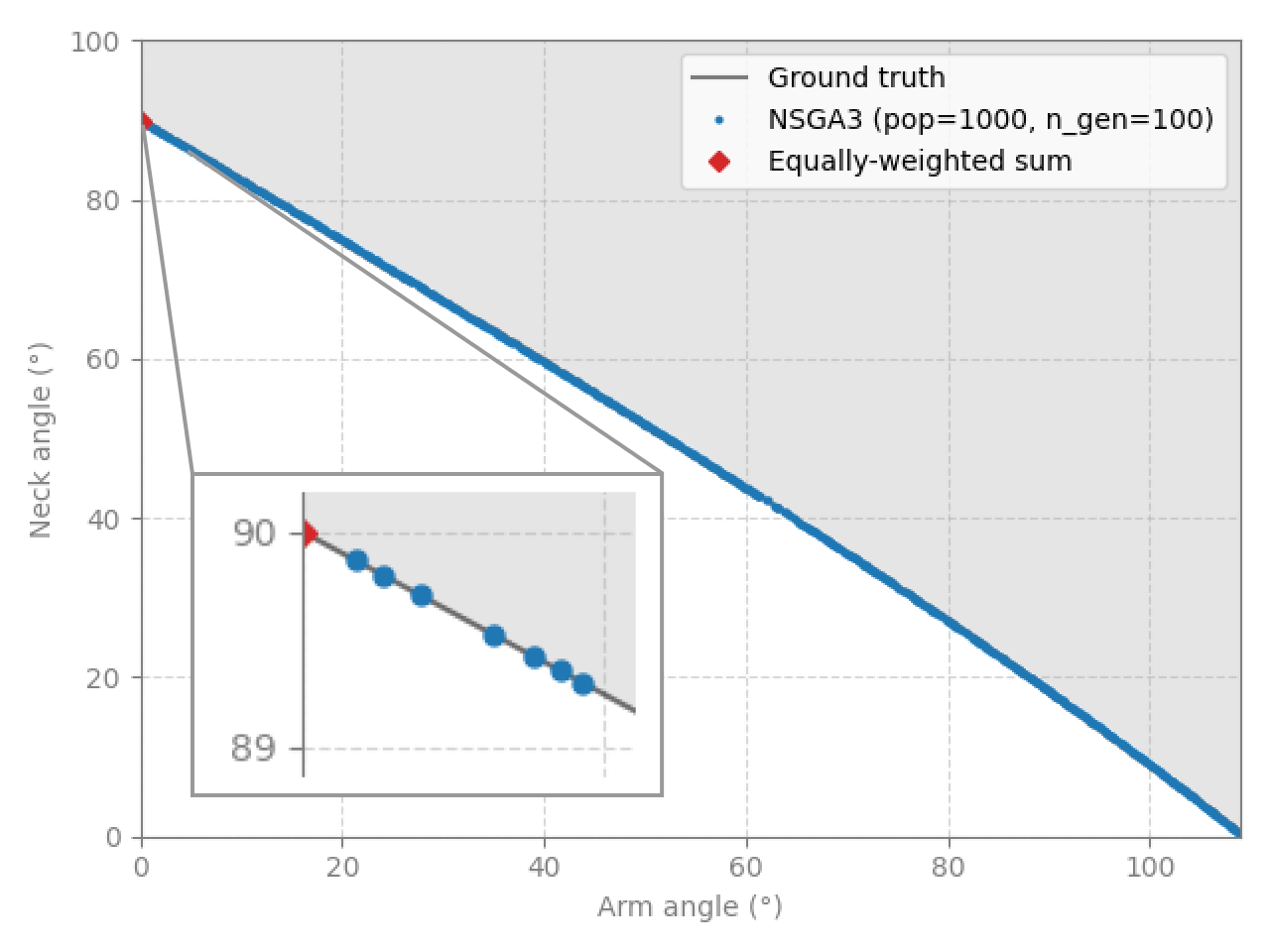}
          \caption{Pareto Frontier}
          \label{fig:pareto_plot}
          \Description{2D scatter plot of a concave Pareto frontier.}
    \end{subfigure}
     \hfill
     \begin{subfigure}[b]{0.4\textwidth}
         \centering
          \includegraphics[width=\linewidth]{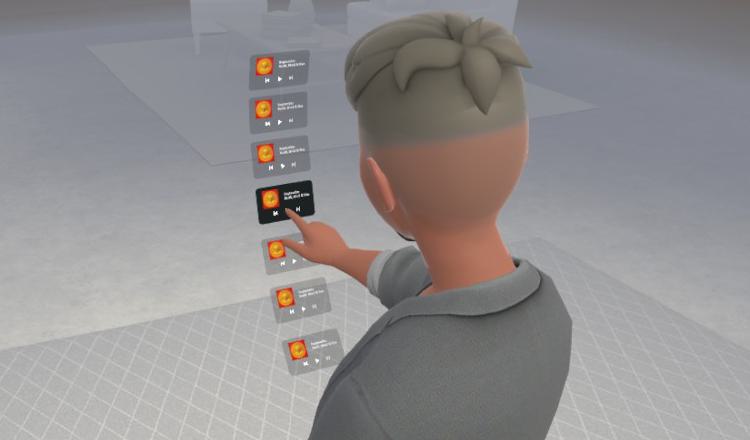}
          \vspace{8pt}
          \caption{3D Scene}
          \label{fig:3D_scene}
          \Description{3D scene with a view over the shoulder of an avatar.}
    \end{subfigure}
     \caption{Results of experimental simulations exploring the effect of optimization technique on adaptations derived via online multi-objective optimization. The simulation is visualized as a 2D illustration of the UI adaptations (left), in a scatter plot containing the true Pareto frontier and the adaptations suggested by the solver (middle), and in a 3D scene featuring the UI (right).}
     \label{fig:experiments}
\end{figure*}

We investigate the potential of our proposed approach by presenting a simulation that explores the impact of the choice of multi-objective optimization techniques on the generated adaptation solutions.
We compare the adaptations suggested by an equally weighted sum with those suggested by a system implementing our proposed approach using the NSGA-III algorithm~\cite{deb_investigating_2019}.
The simulation is based on a simple 3D environment with a single floating UI element (see \autoref{fig:3D_scene}).
The adaptation problem is defined as a constrained bi-objective optimization with the following objectives:

\begin{itemize}
    \item \textbf{Neck angle}: The angle between the user's line of sight in a neutral position (i.e., a vector pointing forward from the user's eyes) and the line connecting the user's head and the UI element (i.e., a vector representing eye gaze toward the UI) should be minimized (see \autoref{fig:illustration}, orange).
    \item \textbf{Arm angle}: The angle between the user's resting arm position and the line connecting the user's shoulder and the UI element should be minimized (see \autoref{fig:illustration}, green).
\end{itemize}

The optimization is constrained by the user's arm length (i.e., all adaptations must be within the user's reach).
In practice, multi-objective UI optimization is likely dependent on a range of usability criteria and constraints \cite{belo_auit_2022}.
For the sake of simplicity and to be able to validate our approach, we have limited our experiments to these problem definitions.
The adaptation system is implemented using a Python-based multi-objective optimization solver utilizing the pymoo package~\cite{blank_pymoo_2020} that can be interfaced with MR adaptation tools such as AUIT~\cite{belo_auit_2022}.
All experiments are conducted on a MacBook Pro with an Apple M1 Pro CPU.

The results show that the equally weighted sum optimization does not yield a fair compromise between the two objectives (see \autoref{fig:experiments}).
It, instead, produces an adaptation equivalent to a single-objective optimum for the arm angle objective due to the concave shape of the Pareto frontier.
While this outcome may be viewed as a result of insufficient normalization, it is not straightforward how to normalize the objective functions in a way that would produce a fair compromise and reflects the inability of weighted sum optimization to retrieve non-convex solutions \cite{marler_survey_2004}.
By contrast, the NSGA-III algorithm accurately approximates the true Pareto frontier --- including the expected Pareto optimal solution that equally balances both objectives ---, mirroring results from operations research literature (e.g., \cite{deb_investigating_2019}).
This suggests that optimization methods that return a representative approximation of the true Pareto frontier, like the NSGA-III algorithm, may form a more suitable basis for UI adaptation in MR applications than the common weighted sum-based optimization.

\section{Discussion}
\label{sec:discussion}

Our experiment indicates that multi-objective optimization algorithms like NSGA-III can find a range of Pareto optimal solutions for relevant and realistic adaptation objectives, whereas weighted sum optimization may produce unexpected or undesired adaptations, depending on objective choice and scaling.
By posing the problem of real-time adaptation of MR UIs as an online multi-objective optimization problem, we have illustrated that it is possible to generate a set of Pareto optimal potential adaptations for MR UI layouts that can be presented to users for selection.
This positions our proposed approach between fully automated techniques, for example, based on global criteria, and fully manual adaptations as present in many current MR systems.
However, our research is still in its early stages and there is a lot of room for further investigation.

% Presentation and Selection
\textit{Presentation and Selection.} It is unclear how and when the potential adaptations should be presented to users, how many solutions to present, and how users should be informed about the trade-offs between different objectives.
It is also uncertain how users should select from the presented adaptations.
We expect large benefits for system usability if the advantages of fully automated and fully manual adaptation can be combined in an adaptation technique.
Techniques such as preference ordering (e.g., calculating distance to a utopia point \cite{marler_survey_2004}) may aid in this issue.

% Cognitive Load
\textit{Cognitive Load.} It is unclear how our approach affects cognitive load, including the presentation and selection techniques and the choice of algorithm, objective functions, and decomposition techniques.
Our proposed interaction might increase cognitive load compared to current automatic single adaptations because it presents multiple adaptation options to the user and allows for more control over the adaptation process. % at the cost of additional options and visual affordances.
It could, however, also decrease overall cognitive load as the user might be able to select and interact with more usable adaptations.
% We currently suspect that our approach will decrease cognitive load as the user gains control over the adaptation.
% We expect that the choice of presentation technique and the current usage context will have a strong influence on overall cognitive load and plan to investigate this effect with end-users.

% Preference Approximation
\textit{Preference Approximation.} How the user's latent preference function can best be approximated and integrated into the optimization procedure remains an open question.
We suspect that multi-objective Bayesian optimization that uses information gain to guide solution set reduction could be a promising candidate, but many alternatives exist.

\section{Conclusion}
\label{sec:conclusion}

In this paper, we proposed a novel technique for real-time UI adaptation in MR applications.
Our approach generates a set of Pareto-optimal potential adaptations that users can choose from, rather than relying on global criterion optimization methods that may not meet their expectations.
We investigated and discussed factors that can impact adaptation usability, including objective choice and scaling and optimization techniques.
Our experiment indicated that multi-objective optimization algorithms like NSGA-III can find a range of Pareto-optimal adaptations for realistic objectives.
This set of potential adaptations can then be further reduced using various decomposition techniques to afford fully automated, semi-automated, or largely manual adaptation.
We have discussed several opportunities for future research, including exploring the presentation and selection of adaptations and evaluating the approach with end-users.
While this work has focused on adapting the position of a single UI element, we suspect that our approach can generalize to multiple-UI scenarios as well as to other problems beyond layout adaptation.
These use cases will require new investigation of presentation and selection techniques and can offer a wealth of avenues for future research.
Overall, we argue that, in light of incomplete and imperfect optimization objectives, adaptive user interfaces should acknowledge the limitations of their optimization methods and emphasize the importance of user choice and control to achieve effective adaptations.
Our approach is one attempt to follow this principle, but many open questions remain that we hope will be explored in future work.

% \begin{acks}
% This paper was supported by...
% \end{acks}

\bibliographystyle{ACM-Reference-Format}
\bibliography{references, references2}

% \appendix

% \input{sections/appendix}

\end{document}